\documentstyle[prb,aps,graphicx]{revtex}
\begin{document}
\pagestyle{empty}

\twocolumn[\hsize\textwidth\columnwidth\hsize\csname
@twocolumnfalse\endcsname
\title{Frustration-Induced Two Dimensional Quantum Disordered Phase in
Piperazinium
Hexachlorodicuprate}

\author{M. B. Stone$^{1}$, I. Zaliznyak$^{1,3}$, Daniel H.
Reich$^{1}$,
and C. Broholm$^{1,2}$}

\address{
$^{1}$Department of Physics and Astronomy, The  Johns Hopkins
University,
Baltimore, MD 21218
\\
$^{2}$National Institute of Standards and Technology, Gaithersburg, MD
20899
\\
$^{3}$Department of Physics, Brookhaven National Laboratory, Upton, NY
11973-5000}

\date{\today}
\maketitle
\thispagestyle{empty}
\begin{abstract}

Piperazinium Hexachlorodicuprate (PHCC) is shown to be a frustrated
quasi-two-dimensional quantum Heisenberg antiferromagnet with a gapped
spectrum.  Zero-field inelastic neutron scattering
and susceptibility and specific heat measurements as a function of
applied magnetic field are presented.  At $T = 1.5$ K, the magnetic 
excitation
spectrum is dominated by a single propagating mode with a gap, $\Delta
= 1$ meV, and bandwidth of $\approx 1.8$ meV in the $(h0l)$ plane.  
The
mode has no dispersion along the $b^*$ direction indicating that
neighboring ${\bf a}$-${\bf c}$ planes of the triclinic structure are 
magnetically
decoupled.  The heat capacity shows a reduction of the gap as a
function of applied magnetic field in agreement with a singlet-triplet
excitation spectrum.  A field-induced ordered phase is observed in
heat capacity and magnetic susceptibility measurements for magnetic
fields greater than $H_{c1}\approx 7.5$ Tesla.  Analysis of the
neutron scattering data reveals the important exchange interactions
and indicates that some of these are highly frustrated.

\end{abstract}
\narrowtext
\vskip0pc]
\newpage

\section{Introduction}

Among physical systems that display collective macroscopic quantum
phenomena, interacting spin systems are perhaps the most
experimentally informative.  Many qualitatively different model
systems are available, and there are numerous experimental tools
providing access to spatial and temporal correlations on microscopic,
mesoscopic, and macroscopic scales.  Owing to their relative
simplicity, much attention has been devoted to one-dimensional
systems, such as the spin S = 1 Haldane chain, \cite{Broholm98} the
even-leg S=1/2 spin ladders, \cite{Carter96,Eccleston98,matsuda99} and
dimerized (alternating bond) S=1/2
chains. \cite{gxu2000,Garrettprb97,Garrettprl97} However, it has long
been a quest for theorists and experimentalists alike to find
analogous cooperative singlet ground state magnetism in higher
dimensions.

For the two-dimensional S=1/2 quantum Heisenberg antiferromagnet (2D 
QHAFM),
a non-trivial quantum-disordered  spin-gap phase with a singlet
ground state in the vicinity of a quantum-critical point has
been studied extensively  by theorists in relation to the magnetic
properties of layered cuprate superconductors. 
\cite{Chakravarty1988,sachdevbook}
Several models were suggested to fall into this quantum-disordered
 phase, ranging from
simple quantum dimer models \cite{Singh1988}, and a valence bond 
crystal
 favored by frustration on the 2D square lattice
\cite{Ueda1996,Zhitomirsky1996}, to the quantum kagom\'{e} 
antiferromagnet\cite{sachdevkagome,singhkagome,Mambrini2000} and the 
long-sought
spin-liquid resonating valence
bond (RVB) state. \cite{Liang1988,Moessner2001}
While there are numerous materials which,
like the layered cuprates, fall into the renormalized-classical 
region of the
phase diagram of the 2D QHAFM, realizations of quantum disordered 2D 
spin
systems close to the quantum critical point are scarce.

There are higher dimensional coupled spin dimer systems such as the
three-dimensional dimer networks $\rm Cs_{3}Cr_{2}Br_{9}$
\cite{leuenberger84} and $\rm KCuCl_{3}$, \cite{cavadini99} and the
quasi-2D material $\rm BaCuSi_{2}O_{6}$.  \cite{sasago97} However, as
with strongly dimerized linear chains such as $\rm Cu(NO_{3})_{2}\cdot
2.5 D_{2}O$, \cite{gxu2000} correlations in these systems are
dominated by isolated spin pairs.  For different reasons, this is also
the case in the newly discovered two-dimensional frustrated spin
system $\rm SrCu_{2}(BO_{3})_{2}$.\cite{kageyama99} In that material,
triplet excitations are localized because of the frustrating symmetry
of the interactions.\cite{Kageyama00} Here we report the discovery of
a frustration-induced spin-singlet phase with a gapped spectrum
in the metalo-organic compound
piperazinium hexachlorodicuprate (PHCC).  The important distinction
from other $D>1$ singlet systems is that there are uniform extensive 
paths of interaction spanning the quasi-2D plane and the strongly 
correlated spin
cluster in PHCC involves several spin pairs. Even so there is a spin 
gap
in the excitation spectrum, and no magnetic phase transition in the 
absence of a magnetic field.

These results were obtained through an extensive characterization of
PHCC using magnetic susceptibility, specific heat, and inelastic
neutron scattering measurements.  Our findings
correct previous reports that suggest that PHCC
is an alternating spin chain
system. \cite{battaglia88,daoud86,regnault89}
In agreement with previous thermodynamic data, we
find a triplet mode with a spin gap $\Delta = 1$ meV, and a bandwidth
of 1.8 meV. However, rather than the dispersion being confined to the
putative chain direction, $\bf{c}$, we observe dispersion throughout 
the
${\bf a}$-${\bf c}$ plane.  In addition, our analysis of the
wavevector dependent scattering intensity shows that no fewer than
five spin pairs are strongly correlated.  Two of these provide {\em
positive} contributions to the ground state energy, indicating the
presence of frustration.  We also present evidence for a field-induced
magnetic phase transition that provides an interesting example of
quantum critical behavior in a frustrated quasi-two-dimensional spin
system.

\section{Piperazinium Hexachlorodicuprate}

Piperazinium hexachlorodicuprate (PHCC), $\rm
(C_{4}H_{12}N_{2})Cu_{2}Cl_{6}$,
has a triclinic crystal structure with space group $P\bar{1}$,
and room temperature  lattice constants
$a = 7.984(4)$ \AA, $b = 7.054(4)$ \AA, $c = 6.104(3)$ \AA, and
$\alpha = 111.23(8)^{\circ}$, $\beta =
99.95(9)^{\circ}$, and $\gamma = 81.26(7)^{\circ}$.
(We use the nomenclature
of Ref.~\cite{battaglia88}.) The
lattice parameters at temperature
$T=1.5$ K were found to be $a = 7.82(2)$, $b = 6.7(3)$, and $c =
5.8(2)$ \AA.

\begin{figure}
    \centering\includegraphics[scale=0.8]{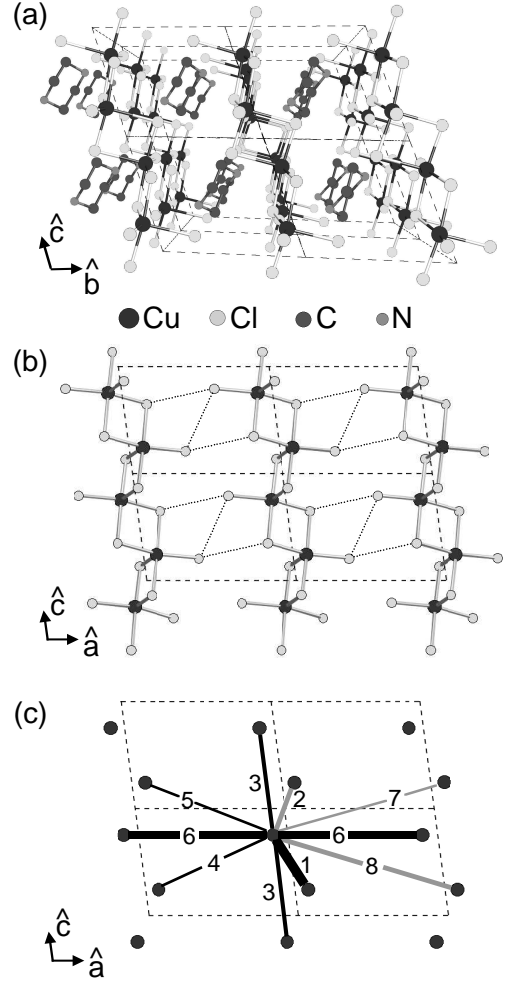}
\caption{\label{fig:phccstruct}
Crystal structure of piperazinium hexachlorodicuprate (PHCC),
$\rm (C_{4}H_{12}N_{2})(Cu_{2}Cl_{6})$.
(a) View along the $\bf{a}$
axis showing well-separated Cu-Cl planes.
(b) A single Cu-Cl plane viewed along the
$\bf{b}$ axis, showing four unit cells.  The dotted lines indicate 
possible
halide-halide contacts (see text).
(c)  Cu-Cu interactions that contribute to
the magnetic Hamiltonian (see Table~\ref{tab:phcctable}.)
The line thickness is proportional to $|J_{\bf d}
\langle {\bf S_0} \cdot {\bf S_d}\rangle |$ for each bond.  The gray 
lines
indicate frustrated bonds. }
\end{figure}

The crystal structure of PHCC is depicted in
Fig.~\ref{fig:phccstruct}.  Figure~\ref{fig:phccstruct}(a) is a
perspective drawing of the crystal structure as viewed along the
$\bf{a}$ axis.  Copper-chlorine layers in the ${\bf a}$-${\bf c}$
plane are clearly visible.  These are well separated from one another
by the piperazinium dication rings.  By rotating the structure in
Fig.~\ref{fig:phccstruct}(a) about the $\bf{c}$ axis, one obtains the
view shown in Fig.~\ref{fig:phccstruct}(b), which depicts a single
copper-chlorine plane.  The $\mathrm{Cu^{2+}}$ ions have a distorted
4+1 coordination with their neighboring chlorine atoms.  The four
short Cu-Cl bonds, with an average bondlength of 2.30 \AA, are
approximately co-planar, with the long 2.62 \AA\ Cu-Cl bond nearly
perpendicular to this plane.  With this coordination, the $\rm
Cu^{2+}$ spin density lies predominantly in the four-bond plane, and
therefore Cu-Cl-Cu superexchange interactions that involve only the
short Cu-Cl bonds should be stronger than those involving the long
bonds.  On this basis, the magnetic Cu-Cu interaction indicated by
Bond 1 in Fig.~\ref{fig:phccstruct}(c) was predicted to be stronger
than Bond 2, \cite{daoud86} leading to the alternating chain model.
\cite{battaglia88,daoud86,regnault89} However, our neutron scattering
results show that while the spins interacting through Bond 1 are
indeed the most strongly correlated, Cu-Cu couplings not previously
considered, such as those due to the halide-halide contacts shown as
dotted lines in Fig.~\ref{fig:phccstruct}(b), result in dispersion
along the $\bf a$ direction that is stronger than that along $\bf c$.
As we will show, PHCC contains a two-dimensional network of magnetic
interactions in the {\bf a}-{\bf c} plane.  The Cu-Cu interactions
that we will consider in our analysis are numbered 1 through 8 in
Fig.~\ref{fig:phccstruct}(c).  Our results pertaining to these bonds
are listed in Table~\ref{tab:phcctable}.

\section{Experimental Techniques}

Powder samples of PHCC were prepared by rapid cooling
from 50 $^{\circ}$C to 0 $^{\circ}$C of
38\% hydrochloric acid solutions
containing piperazinium dihydrochloride and copper(II) chloride in a
1:4 molar ratio.  \cite{battaglia88} Single crystals were grown by
slowly reducing the temperature of similar saturated solutions from
50 $^{\circ}$C to 18 $^{\circ}$C over 75 hrs.  Seed crystals obtained
in this manner were suspended in saturated solutions and grown
further.  The crystals typically grow as dark red tablets with $(100)$
faces providing the largest facets.  Deuterated single crystals for
the inelastic neutron scattering measurements were produced in this
manner from commercially available piperazinium$\mathrm{(d_{8})}$
dihydrochloride and anhydrous copper chloride(II), using 35\% DCl in
D$_{2}$O as the solvent.   The crystals were 89(1)\%
deuterated, as determined by neutron activation analysis.

Low field  DC magnetic susceptibility measurements
were performed on a powder
sample of mass $m = 16$ mg
in the temperature range 1.7 K $< T <$ 270 K using a
SQUID magnetometer.  AC  susceptibility
measurements using a balanced-coil susceptometer
were performed
in the temperature range 0.125 K $< T <$ 8 K
on single crystals of typical mass $m
= 20$ mg, in DC fields up to 9 Tesla
 oriented parallel to the $\bf{c}$ axis.
Specific heat measurements were performed on a single crystal of mass
$m =4$ mg for  0.125 K $ < T <2.5$ K in fields up to 9 Tesla
using relaxation calorimetry \cite{bachmann72} with
the field  oriented  parallel to the $\bf{a^{*}}$ axis.

The sample used for inelastic neutron scattering
measurements consisted of three deuterated
single crystals with a total mass of
$m=3.32$ grams coaligned within $0.8^{\circ}$.
 Measurements in the $(0kl)$ and $(h0l)$ scattering
planes were performed on the SPINS cold neutron triple axis
spectrometer at
the National Institute of Standards and Technology (NIST)
Center for Neutron Research
in Gaithersburg, MD.  The horizontal
beam collimation before the sample was $50^{\prime}/k_{i}$
(\AA$^{-1}$)
- 80$^{\prime}$ for the $(0kl)$
measurements, and  $50^{\prime}/k_{i}$ (\AA$^{-1}$)
- 76$^{\prime}$ for the $(h0l)$ measurements. A liquid
nitrogen cooled BeO filter was placed after the sample, and data
were collected at fixed final energy $E_{f} = 3.7$ meV.  A
horizontally focusing pyrolytic graphite (PG(002)) analyzer with
acceptance
angles of 2.7$^o\times$7.2$^o$ in the horizontal and vertical plane
respectively
was used to increase the count rate
at the expense of broadening the instrumental wavevector resolution
perpendicular to
the scattered neutron wavevector,  ${\bf  k}_{f}$.
At energy transfer $\hbar\omega = 0$,
the energy resolution
was $\delta \hbar \omega = 0.13$ meV, and
the instrumental resolution ellipsoid \cite{ChesserAxe}
projected
on the scattering plane had
Full Width at Half
Maximum (FWHM) principal axes
$\delta Q_{1} = 0.043$ \AA$^{-1}$ and
$\delta Q_{2} = 0.076$ \AA$^{-1}$.

Measurements in the $(hk0)$ plane were performed on the BT2 thermal
neutron triple axis spectrometer at NIST.  Horizontal collimations of
$60'-20'-20'-60'$ were used, and data were collected
at $E_{f} = 13.7$ meV with a PG filter before the analyzer.
This gave typical FWHM resolutions
at $\hbar\omega = 0$
of $\delta \hbar \omega =  0.84$ meV, and
 $\delta Q_{\|} =
0.02$\AA$^{-1}$ and  $\delta Q_{\bot} = 0.09$\AA$^{-1}$
parallel and perpendicular to ${\bf Q}$, respectively.
All data were converted to the normalized
scattering intensity $\tilde{I}({\bf Q}, \hbar \omega)$
following a procedure detailed elsewhere, \cite{hammar98}
using the incoherent elastic scattering  of the sample as measured
in each experimental configuration.

\section{Experimental Results}

\subsection{Magnetic Susceptibility}
\begin{figure}
    \centering\includegraphics[scale=0.75]{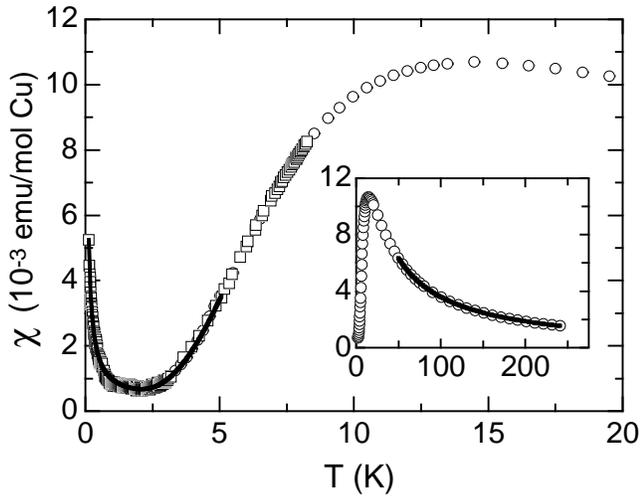}
\caption{\label{fig:chivst}
Magnetic susceptibility $\chi(T)$ of PHCC.  Open circles: DC
suceptibility of powder.  Open squares: AC susceptibility of
single crystal.  Inset: solid line
is a fit to a Curie-Weiss law.  Main frame: solid
 line is a fit to the asymptotic low-temperature susceptibility
of a 2D gapped system as described in
the text.}
\end{figure}

The DC magnetic susceptibility $\chi(T) = M/H$ for a powder sample of
PHCC is shown in Fig.~\ref{fig:chivst}.  The data were taken in an
applied field $H = 50$ Oe, and the magnetization $M$ was found to be
linear in field up to $H = 1$ Tesla at $T = 1.8$ K. As shown in the
inset of Fig.~\ref{fig:chivst}, $\chi(T)$ rises with decreasing $T$ to
a rounded maximum at $T \approx 12$ K, followed by a rapid decrease,
as has been previously observed.  \cite{battaglia88,daoud86} The low
temperature behavior of $\chi(T)$ is shown in the main panel of
Fig.~\ref{fig:chivst}, which also includes single-crystal AC
susceptibility data measured down to $T = 0.125$ K. Apart from a
paramagnetic background at the lowest temperatures that is
attributable to residual impurities, these data show the exponentially
activated dependence of $\chi(T)$ characteristic of a gapped
Heisenberg antiferromagnet.

The AC magnetic susceptibility for a single crystal
as a function of applied magnetic field
at fixed temperature is shown in Fig.~\ref{fig:chivsh}.
The sharp feature in the data indicates an abrupt change in the
magnetic density of states of PHCC as
the spin gap closes at the critical field $H_{C1} \approx 7.5$ T
at $T = 0.125$ K.
This feature moves to larger magnetic field as the
temperature is increased, and eventually becomes
smeared out due to thermal population of
multiple energy levels above the spin gap.

\subsection{Specific Heat}
\begin{figure}
    \centering\includegraphics[scale=0.5]{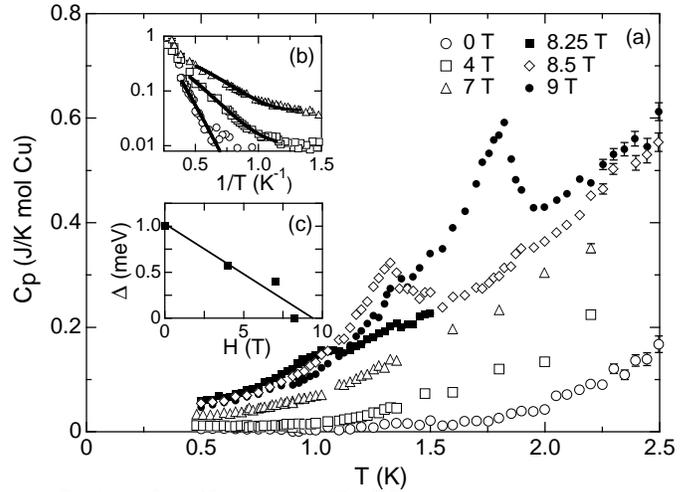}
\caption{\label{fig:cpvst}
Specific heat of PHCC versus temperature at constant applied magnetic
field.
(b) shows activated behavior at low fields and fits used to
determine field-dependence of the spin gap shown in (c). }
\end{figure}

Figure~\ref{fig:cpvst} depicts the heat capacity as a function of
temperature measured for different fixed magnetic fields up to $H =9$
T. The lattice contribution to the heat capacity has not been
subtracted from the data.  The exponentially activated heat capacity,
which is apparent from the inset to Fig.~\ref{fig:cpvst} for fields $H
<8$ Tesla, is further direct evidence for a spin gap in PHCC. As $H$ 
is
increased, there is an increase in the specific heat at lower
temperatures and the activation energy decreases, effects that
indicate that the spin gap is decreasing.  For $H > 8$ T, there is a
singularity in the temperature dependence of the specific heat
indicative of a phase transition.  This implies that above
$H_{C1}$ PHCC enters a 3D-ordered phase, as is typical of quasi-1D and
quasi-2D gapped quantum antiferromagnets. Both the transition
temperature and the the spin entropy
removed from the system via the transition  increase with $H$.

\subsection{Inelastic Neutron Scattering}
\begin{figure}
    \centering\includegraphics[scale=0.5]{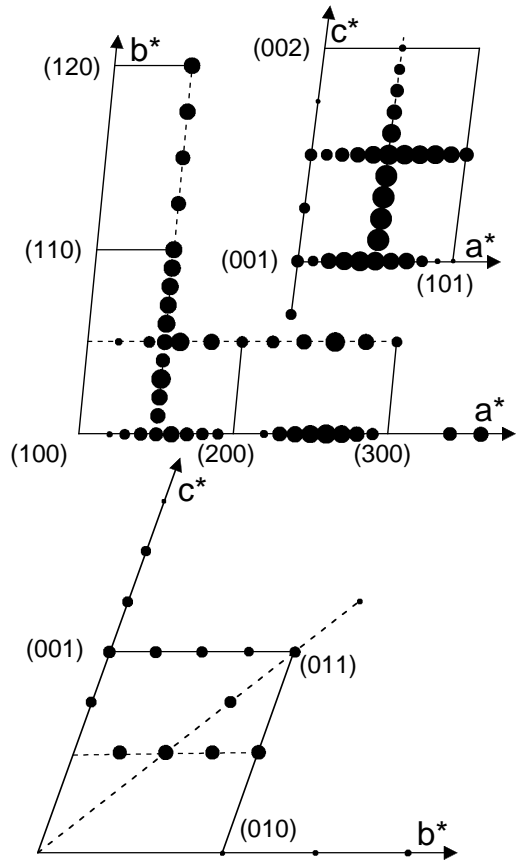}
\caption{\label{fig:phccscans}
Points in reciprocal space measured using inelastic neutron
scattering.  The area of each point is proportional
to the measured first moment of the spectrum at that
{\bf Q}.}
\end{figure}

\begin{figure}
    \centering\includegraphics[scale=0.65]{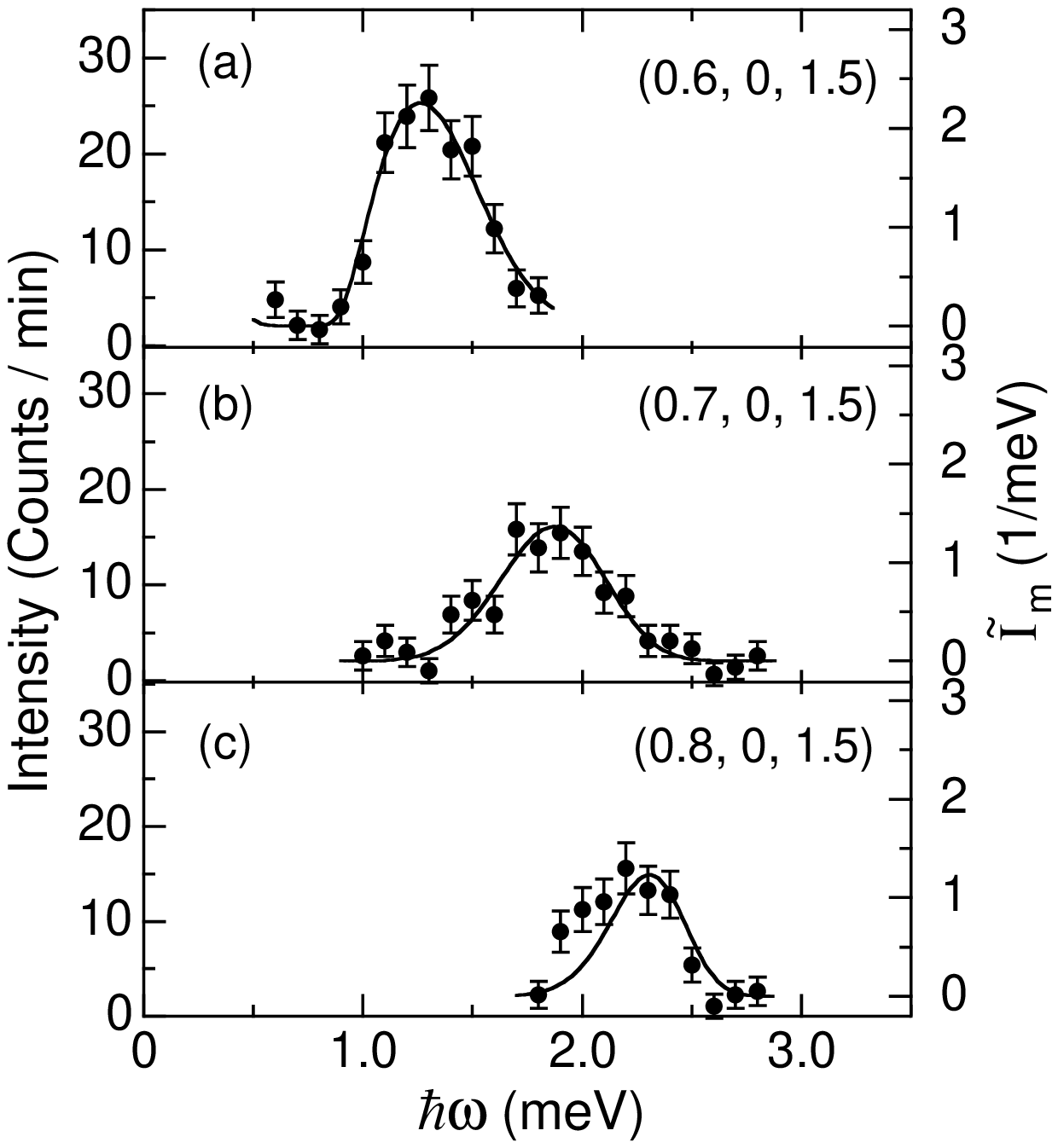}
\caption{\label{fig:escansH}
Inelastic neutron scattering data for PHCC at $T = 1.5 $ K
showing dispersion
of magnetic excitation with $h$.  Right axis shows normalized
magnetic scattering intensity $\tilde{I}_{m}({\bf Q},\omega)$.
The solid lines are a fit to a two-dimensional  model described
in the text.}
\end{figure}

\begin{figure}
    \centering\includegraphics[scale=0.65]{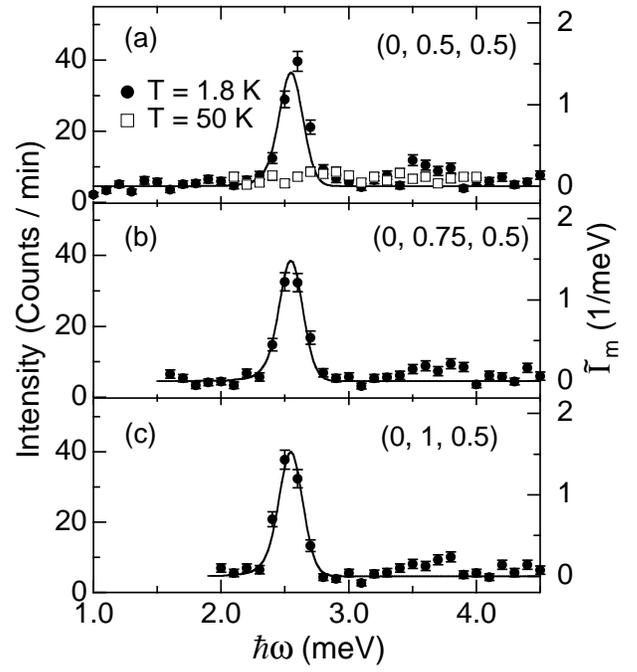}
\caption{\label{fig:escansK}
Inelastic neutron scattering data for PHCC at $T = 1.5 $ K showing
lack of dispersion
of magnetic excitation with $k$. Open symbols in (a): data
taken at $T = 50$ K. Right axis shows normalized
magnetic scattering intensity $\tilde{I}_{m}({\bf Q},\omega)$.
The solid lines are a fit to a two-dimensional  model described
in the text.}
\end{figure}

\begin{figure}
    \centering\includegraphics[scale=0.65]{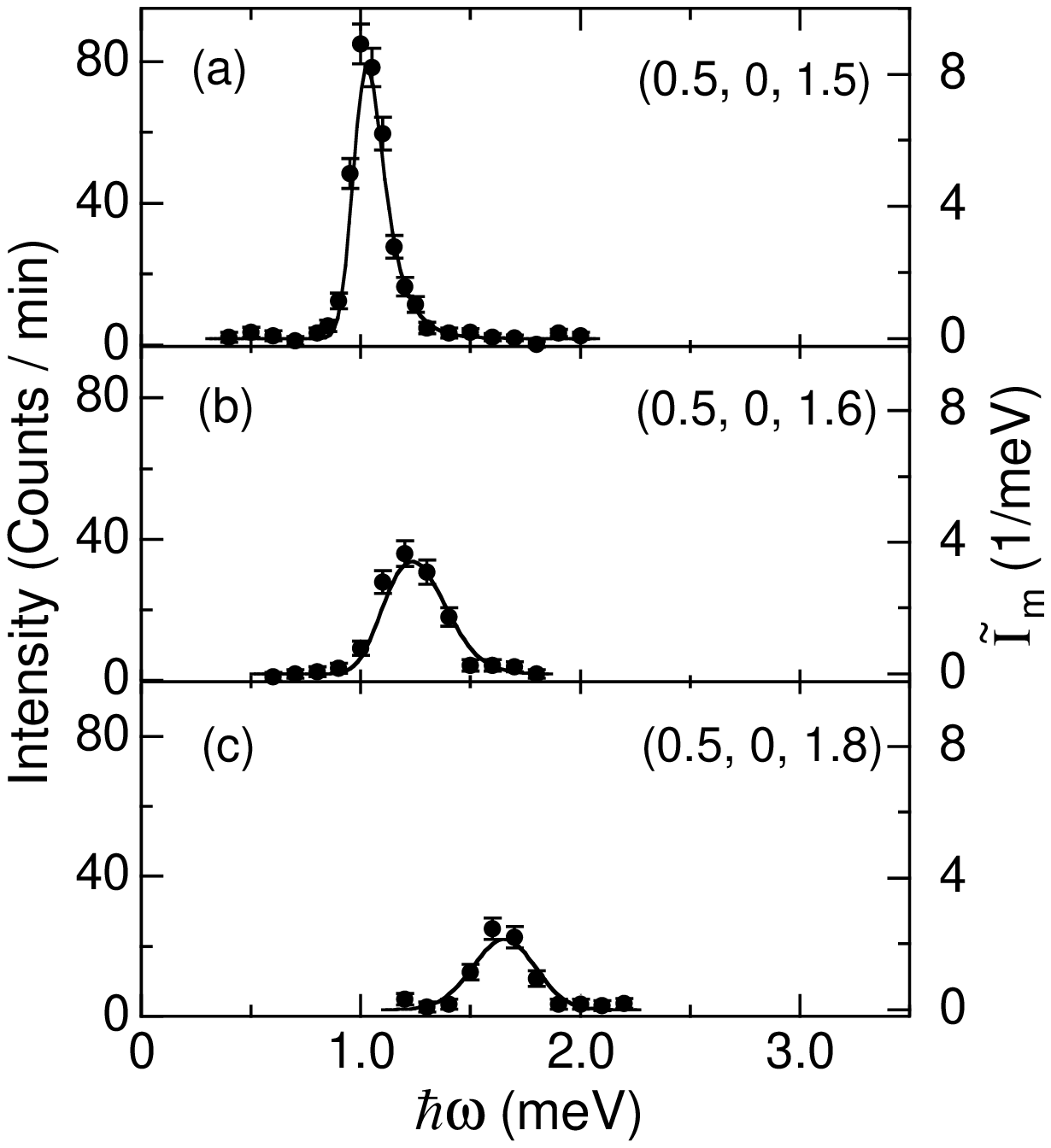}
\caption{\label{fig:escansL}
Inelastic neutron scattering data for PHCC at $T = 1.5 $ K
showing dispersion
of magnetic excitation with $l$. Right axis shows normalized
magnetic scattering intensity $\tilde{I}_{m}({\bf Q},\omega)$.
The solid lines are a fit to a two-dimensional  model described
in the text.}
\end{figure}

\begin{figure}
    \centering\includegraphics[scale=0.7]{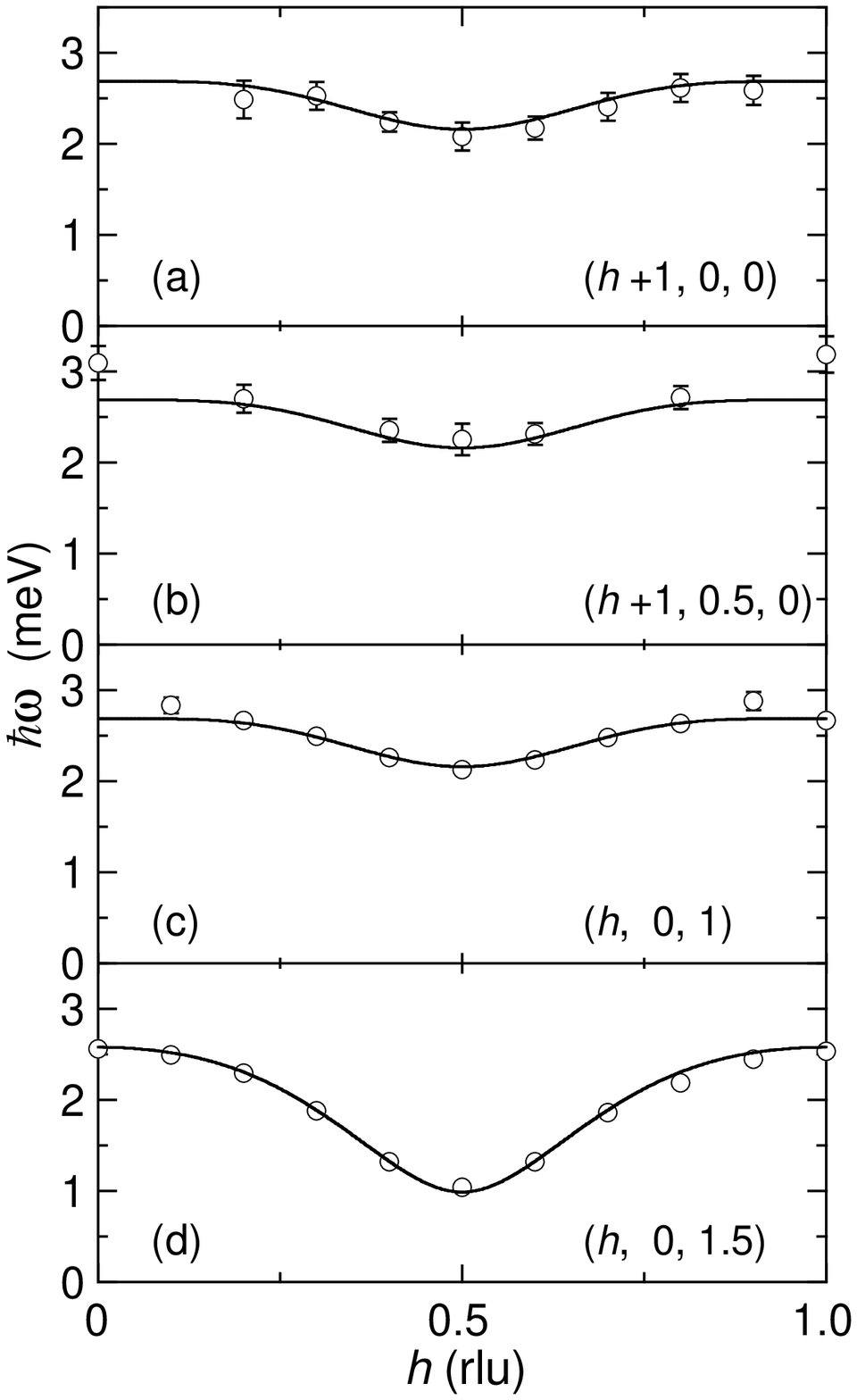}
\caption{\label{fig:phccdispH}
Energy of magnetic excitations in PHCC
showing dispersion with $h$ at constant $k$ and $l$.  Data
points are determined from Gaussian fits to constant-{\bf Q} scans.
The solid lines are a fit to a two-dimensional  model described
in the text.}
\end{figure}

\begin{figure}
    \centering\includegraphics[scale=0.7]{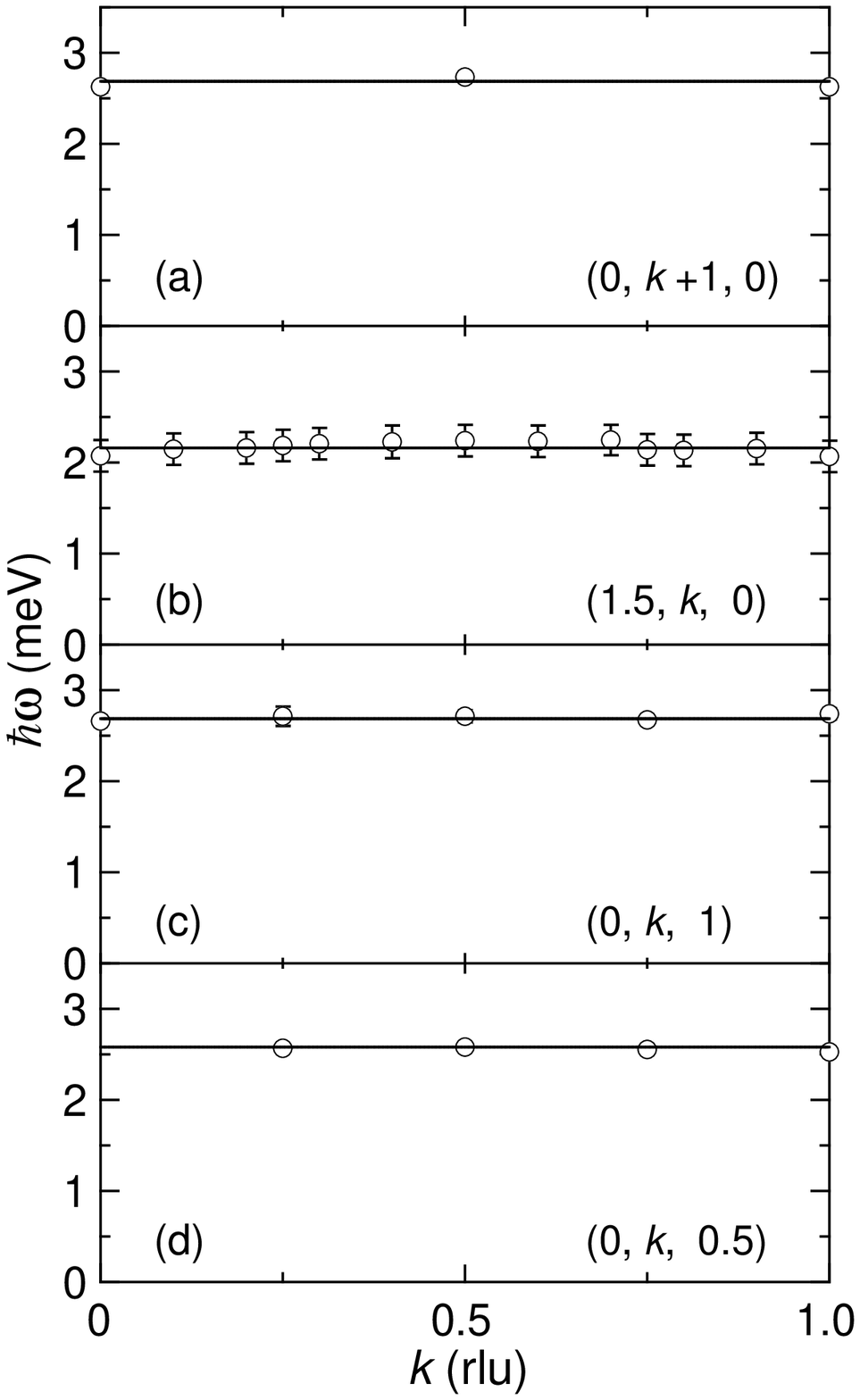}
\caption{\label{fig:phccdispK}
Energy of magnetic excitations in PHCC
showing lack of dispersion with $k$ at constant $h$ and $l$.  Data
points are determined from Gaussian fits to constant-{\bf Q} scans.
The solid lines are fit to a two-dimensional  model described
in the text.}
\end{figure}

\begin{figure}
    \centering\includegraphics[scale=0.7]{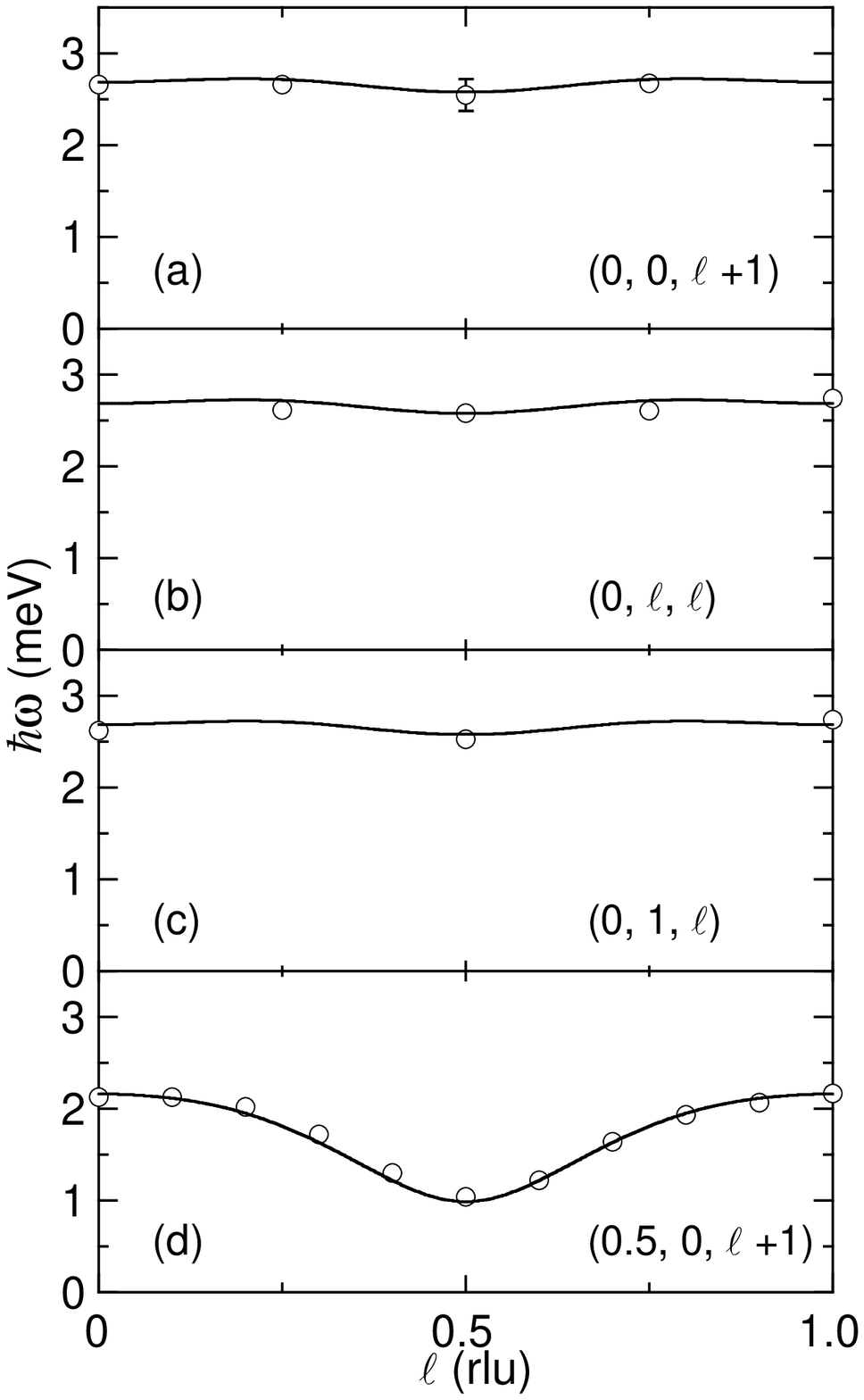}
\caption{\label{fig:phccdispL}
Energy of magnetic excitations in PHCC
showing dispersion with $l$ at constant $h$ and $k$.  Data
points are determined from Gaussian fits to constant-{\bf Q} scans.
The solid lines are fit to a two-dimensional  model described
in the text.}
\end{figure}

To map out the wavevector dependence of the magnetic
excitation spectrum of PHCC,
energy transfer scans at constant {\bf Q}
were performed at 93 different locations
throughout the reciprocal space of PHCC in zero magnetic
field at $T = 1.5$ K.
Figure~\ref{fig:phccscans}
depicts the three reciprocal lattice planes and
the particular points in reciprocal space that were probed.
At every wavevector measured, the spectrum is dominated
by a single, resolution-limited mode.
Representative energy scans are shown in
Figs.~\ref{fig:escansH}, ~\ref{fig:escansK}, and ~\ref{fig:escansL}.
They illustrate dispersion of this mode along the $h$ and
$l$ directions and the absence of dispersion as wavevector transfer
varies along the $k$ direction.
The mode disappears upon heating to T=50 K, as shown in 
Fig.~\ref{fig:escansK}(a),
demonstrating that the observed scattering is magnetic in origin.
The mode has its minimum energy (spin gap) $\Delta = 1$ meV
at ${\bf Q} = (0.5,0,1.5)$ [Fig.~\ref{fig:escansL}(a)].
The full  dispersion of this mode is summarized in
Figs.~\ref{fig:phccdispH}, \ref{fig:phccdispK} and
\ref{fig:phccdispL}.
Each point in these figures was obtained by fitting a scan such
as those in Figs.~\ref{fig:escansH}, ~\ref{fig:escansK}, and
~\ref{fig:escansL}
to a Gaussian lineshape.
From these figures, the two-dimensional nature of the magnetic
interactions of PHCC becomes clear.
Figs.~\ref{fig:phccdispH}(d) and \ref{fig:phccdispL}(d) show
significant dispersion along lines in
reciprocal space that intersect the two-dimensional magnetic zone
center
at ${\bf Q} = (0.5,0,1.5)$, with somewhat stronger dispersion
along $h$ than along $l$.  Along the edges of the magnetic
zone, additional dispersion with
$h$ is seen in Figs.~\ref{fig:phccdispH}(a)-(c),
but only minimal variation in peak position was observed
with $l$ in Figs.~\ref{fig:phccdispL}(a)-(c).
From Fig.~\ref{fig:phccdispK}, an upper bound of
0.2(2) meV can be set on variation of
the mode energy with $k$.
The solid curves in
Figs.~\ref{fig:escansH}-\ref{fig:phccdispL} result from a fit using a
two-dimensional model for the dynamic
spin correlation function as described below.

\section{Discussion}

\subsection{Magnetic Susceptibility}

The zero field susceptibility data yield significant information
about the microscopic spin Hamiltonian.  First, a measure
of the size of the spin gap may be obtained.
In $D$ dimensions, the asymptotic low-temperature
susceptibility of a gapped spin system
with quadratic dispersion \cite{Troyer94} is proportional to
\begin{equation}
\chi(T) \propto T^{(D/2) -1}e^{-\Delta/k_{B}T} .
\label{eq:chigapped}
\end{equation}
The  solid line in  the main panel of Fig.~\ref{fig:chivst} is a fit 
to
the two-dimensional form of Eq.~\ref{eq:chigapped}, including a
diamagnetic
background and a low-temperature Curie tail.
The fit was restricted to $T \le 5$ K, and gave a spin gap
 $\Delta =  1.2(1)$
meV, in good agreement with that observed in the neutron scattering
data.

From the high temperature Curie-Weiss behavior of $\chi (T)$ we
obtain
a value for the sum of all
relevant exchange constants in the system
\begin{equation}
J_{0}=\sum_{{\bf d}}J_{{\bf d}} ,
\end{equation}
where $\{{\bf d}\}$ is the set of bonds connecting a spin to its
neighbors.
Fitting the sum of a diamagnetic and a Curie-Weiss term to our
$\chi(T)$  data
in the temperature range 50 K $ < T < 250 $ K we extracted the
Curie-Weiss temperature, $\Theta_{CW}$, the average $g-$factor and
the  diamagnetic term resulting from the sample and sample
holder. The fit is shown as a solid line in the inset to
Fig.~\ref{fig:chivst}. The   Curie-Weiss temperature
$\Theta_{CW} = -19.0(7)$ K is related to $J_{0}$ through
$\Theta _{CW} = \frac{S(S+1)J_{0}}{3 k_{B}}$. \cite{ashcroft} Solving
for
the sum of the independent exchange constants, we
find $J_{0} = 3.3(1)$ meV while
$g_{av} = 2.2(3)$.  A similar estimate
for $J_{0}$ is obtained by fitting $\chi(T)$ to an alternating
spin chain model. \cite{battaglia88,daoud86} This reflects
the insensitivity of the zero-field susceptibility to the geometry
of the magnetic interactions.

 \subsection{Specific Heat}
The change in the activated temperature dependence of the
specific heat with increasing magnetic field for
$H < 8$ T gives a measure of the reduction of the spin
gap as the lower critical field is approached.
The asymptotic low-temperature form for the specific heat, derived
under the same conditions as for Eq.~\ref{eq:chigapped} is given by
\begin{equation}
C_{p}(T) \propto T^{(D/2) -2}e^{-\Delta/k_{B}T} .
\label{eq:cpgapped}
\end{equation}
The solid lines in  Fig.~\ref{fig:cpvst}(b)
are fits to this form for $D =2$,    with the
addition of a term proportional
to $H^{2}/T^{2}$ to account for nuclear
spin contributions at the lowest temperatures.
The derived gap values are shown
versus the applied field in Fig.~\ref{fig:cpvst}(c).
The approximately linear reduction of the spin gap with field
is consistent with the Zeeman splitting of a triplet of excited 
states.

\begin{figure}
    \centering\includegraphics[scale=0.5]{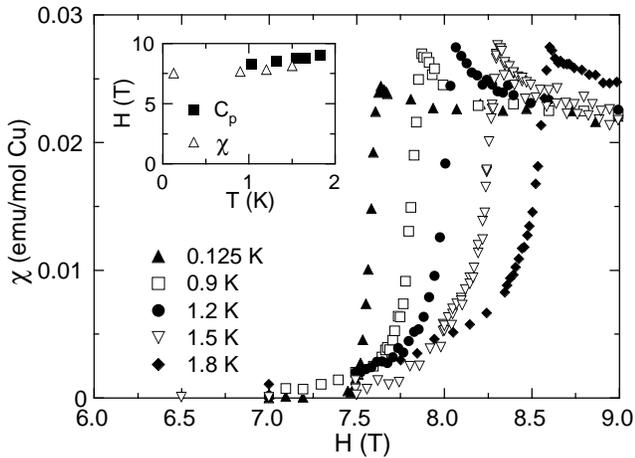}
\caption{\label{fig:chivsh}
AC magnetic susceptibility $\chi(H)$ of single crystal PHCC at
constant temperatures  $T=0.125$, 0.9, 1.2, 1.5, and 1.8 K.
 Inset:  Portion of $H-T$ phase diagram of PHCC derived from
specific heat (filled squares) and
magnetic susceptibility (triangles) data.}
\end{figure}

The positions of the peaks observed in the temperature dependent 
specific
heat
for $H > 8$ T outline the lower edge of the ordered region
of the $H$-$T$ phase diagram. This phase boundary is shown
in the inset to Fig.~\ref{fig:chivsh}. A rough
estimate for the upper critical field
is $H_{C2} = J_{0}/g_{av}\mu_{B} = 26(3) $T.
Clearly, further work is required to map out the full extent of
the ordered
phase and also to determine the structure and symmetry of the order
parameter.
The locations of peaks in the
field-dependent, single crystal magnetic susceptibility data are
also shown in the inset to Fig.~\ref{fig:chivsh}.
Taking into account
the anisotropic g-tensor of PHCC\cite{daoud86}, these data are 
entirely
consistent
with the specific heat data.

\subsection{Inelastic Neutron Scattering}

The magnetic contribution, $\tilde{I}_{m}({\bf Q}, \hbar \omega)$,
to the normalized neutron scattering intensity
$\tilde{I}({\bf Q}, \hbar \omega)$
is related to
the dynamic spin correlation function
${\cal S}^{\alpha \beta}({\bf Q},\omega )$ as follows
\cite{hammar98,lovesey}
\begin{eqnarray}
\label{eq:ItildeQw}
 \tilde{I}_m({\bf Q},\hbar\omega) = & \int d^3Q'\hbar
d\omega'{\cal R}_{\bf Q, \omega}({\bf Q}-{\bf Q}',\omega - \omega') \\
\nonumber
 & |\frac{g}{2}F(Q')|^2 \sum_{\alpha \beta} \left(\delta_{\alpha 
\beta}
  - \hat{Q}_\alpha'\hat{Q}_\beta'\right)
 {\cal S}^{\alpha\beta}({\bf Q}',\omega').
\end{eqnarray}
Here $F(Q)$ is the magnetic form factor of the $\mathrm{Cu}^{2+}$
ion, and ${\cal R}_{{\bf Q}\omega }$ is the normalized instrumental
resolution function. \cite{ChesserAxe}
We assume that the magnetic Hamiltonian is
isotropic in spin space, and thus since no symmetry-breaking ordering
transition occurs in PHCC at zero field,
the dynamic spin correlation function should
also be isotropic. This is supported by  previous susceptibility
measurements, which indicate that exchange anisotropy in PHCC is
very small \cite{daoud86}, and by the absence of any observable 
anisotropy splitting of the excitations observed by neutron 
scattering. In this case the different spin
polarizations in Eq.~\ref{eq:ItildeQw}
can be summed  to give

\begin{equation}
\label{Sq}
\sum_{\alpha \beta} \left(\delta_{\alpha \beta}
 - \hat{Q}_\alpha'\hat{Q}_\beta'\right) {\cal S}^{\alpha\beta}
 ({\bf Q}',\omega') = 2{\cal S}^{\alpha\alpha} ({\bf Q}',\omega').
\end{equation}

\begin{figure}
    \centering\includegraphics[scale=0.7]{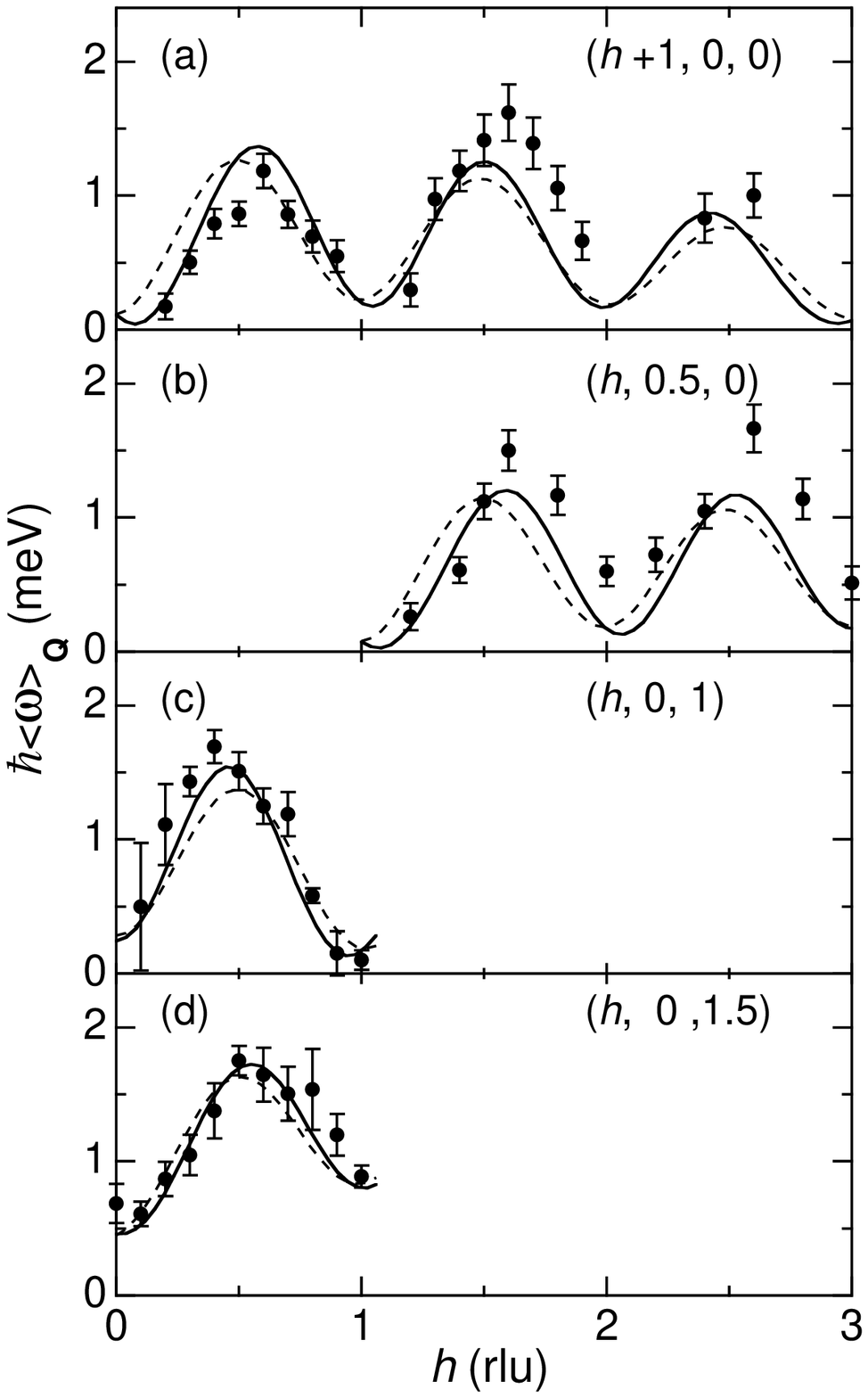}
\caption{\label{fig:fmh}
Variation of first moment
$\hbar \langle \omega\rangle_{\bf Q}$
of PHCC with $h$ at constant $k$ and $l$.
Data points are determined by fitting each original constant-{\bf Q}
scan to a Gaussian
peak and calculating the first moment of the intensity from that
fit.  The dashed(solid) lines are determined by a global fit including
6(8) Cu-Cu interactions, as described in the text.}
\end{figure}

\begin{figure}
    \centering\includegraphics[scale=0.7]{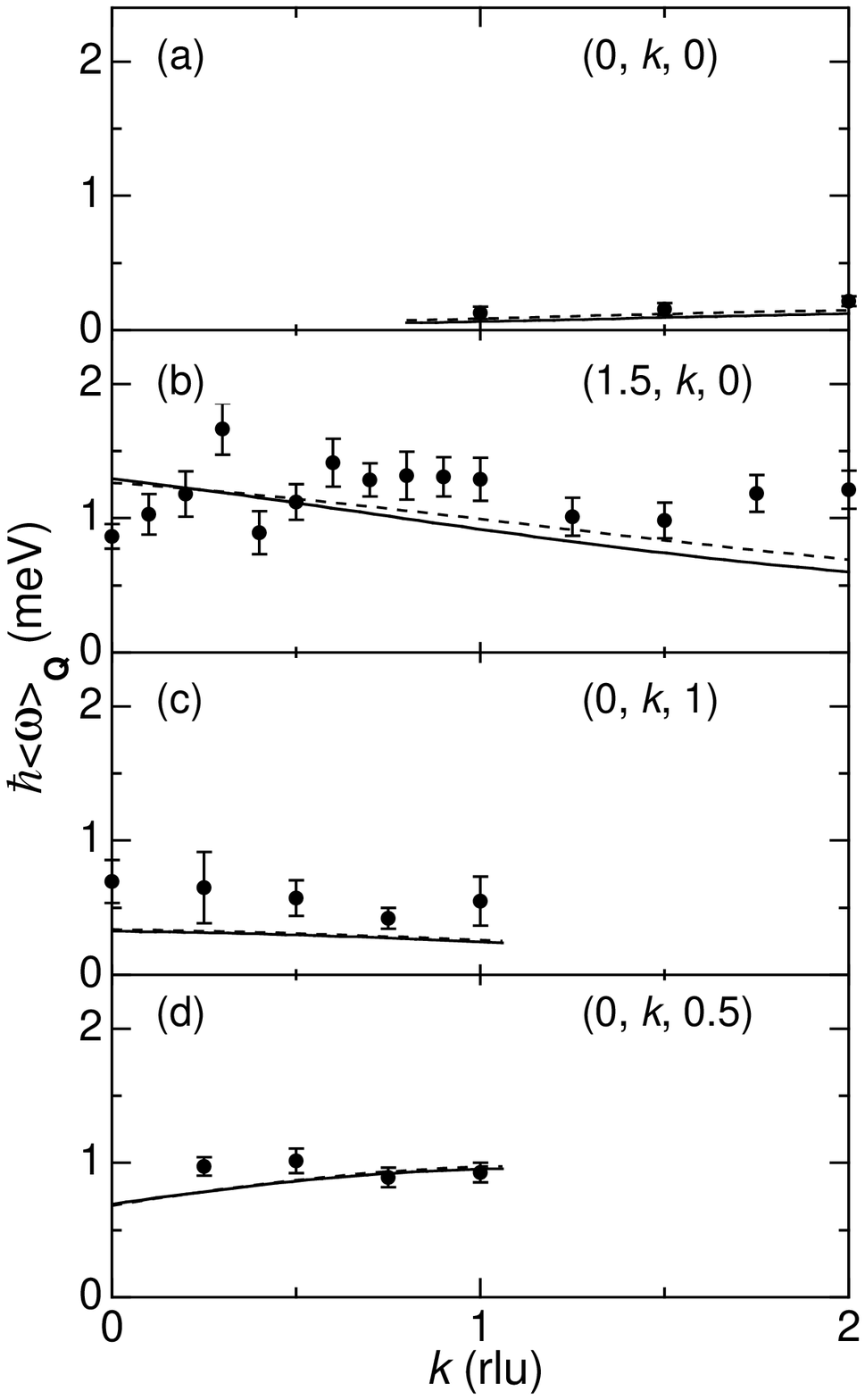}
\caption{\label{fig:fmk}
Variation of first moment
$\hbar \langle \omega\rangle_{\bf Q}$
of PHCC with $k$ at constant $h$ and $l$.
Data points are determined by fitting each original constant-{\bf Q}
scan to a Gaussian
peak and calculating the first moment of the intensity from that
fit.  The dashed(solid) lines are determined by a global fit including
6(8) Cu-Cu interactions, as described in the text.}
\end{figure}

\begin{figure}
    \centering\includegraphics[scale=0.7]{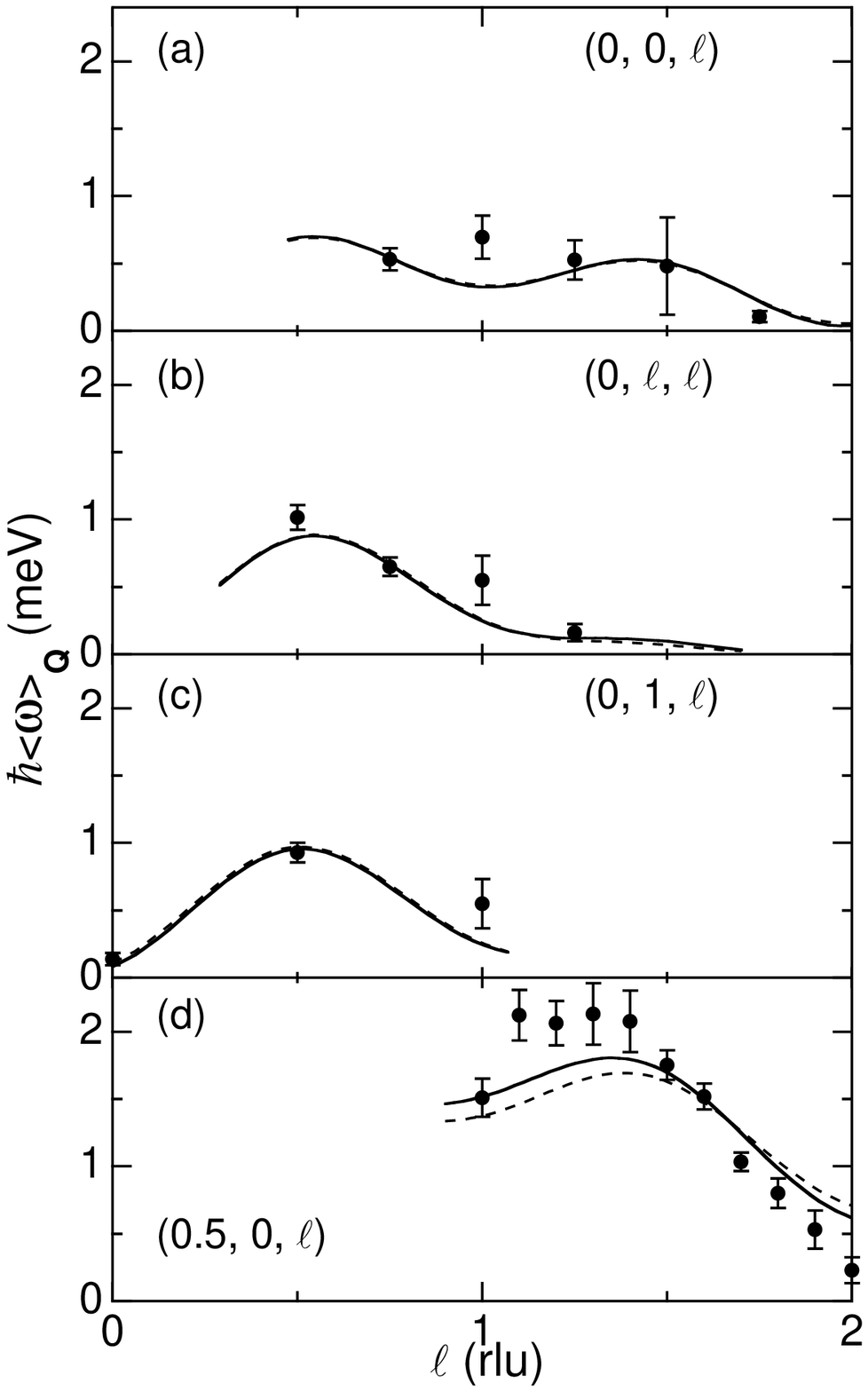}
\caption{\label{fig:fml}
Variation of first moment
$\hbar \langle \omega\rangle_{\bf Q}$
of PHCC with $l$ at constant $h$ and $k$.
Data points are determined by fitting each original constant-{\bf Q}
scan to a Gaussian
peak and calculating the first moment of the intensity from that
fit.  The dashed(solid) lines are determined by a global fit including
6(8) Cu-Cu interactions, as described in the text.}
\end{figure}

Owing to the low symmetry of PHCC, there are a large number of
potential Cu-Cu magnetic interactions within the $\bf a-c$
plane. Information about the relative importance of spin pair
correlations to the ground state energy can be obtained in a model
independent way through the first moment sum rule. \cite{hohenberg74}
For an isotropic spin system, this is
\begin{eqnarray}
\hbar \langle \omega\rangle_{\bf Q}&\equiv&
\hbar^2\int_{-\infty}^{\infty}\omega  {\cal S}^{\alpha \alpha}({\bf
Q},\omega
)d\omega \nonumber \\
&=& -\frac{1}{3}
\sum_{\bf d}J_{\bf d}\langle{\bf S}_0\cdot{\bf S}_{\bf d}\rangle
(1-\cos {\bf Q\cdot d}) ,
\label{eq:firstmomsumr}
\end{eqnarray}
where $J_{\bf d}$ is the
exchange strength and $\langle{\bf S}_0\cdot{\bf S}_{\bf d}\rangle$ 
is the
two-spin correlation function for the spin-pair with bond vector
${\bf d}$. We write
the Hamiltonian in the form
\begin{equation}
{\cal H}= \frac{1}{2}\sum_{\bf r,d} J_{\bf d}{\bf S}_{\bf r}\cdot
{\bf S}_{\bf r+d} ,
\end{equation}
where the index $\{{\bf r}\}$  runs over all spins.
The ground state energy per spin $E_{0}$ is
closely related to the first moment, being simply
\begin{equation}
E_{0} = \frac{1}{2}\sum_{\bf d}J_{\bf d}\langle{\bf S}_0\cdot{\bf 
S}_{\bf
d}\rangle  .
\label{eq:e0}
\end{equation}
The first frequency moment of the dynamic correlation function
$\hbar\langle \omega \rangle_{\bf Q}$ measured in PHCC is shown in
Fig.~\ref{fig:phccscans} and
Figs.~\ref{fig:fmh}-\ref{fig:fml}.
The data points were obtained from the same Gaussian fits
that were used to determine the mode energy.

Unlike what is observed in strongly dimerized
spin gap systems,  the  variation of
$\hbar \langle \omega\rangle_{\bf Q}$ with ${\bf Q}$
in PHCC cannot be accounted for by a single spin pair. Instead,
several crystallographically distinct spin pairs are strongly
correlated and contribute significantly to the ground state energy.
This renders the system
considerably more intricate to describe theoretically.
For example, the RPA theory based on a dimerized
ground state that accounts for strongly dimerized systems in one, two,
and three
dimensions\cite{matsuda99,gxu2000,leuenberger84,cavadini99,sasago97}
is not adequate here since a single spin pair controls the first 
moment of
${\cal S}^{\alpha \alpha}({\bf Q},\omega )$ in that model.

In the absence of an adequate theory of gapped spin systems for $D > 
1$
that goes beyond a dimer-based expansion, we parametrize the
measured dispersion $E({\bf Q})$ in PHCC with the following
phenomenological expression, consistent with Bloch's theorem:
\begin{eqnarray}
E({\bf Q}) &=&   ( B_{0} +B_{h}\cos(2 \pi h)
+ B_{l} \cos(2 \pi l) \nonumber \\
&+& B_{hl} [\cos(2\pi(h+l)) +
\cos(2 \pi (h-l))] \nonumber \\
&+& B_{2h}\cos(4 \pi h) + B_{2l} \cos(4 \pi l)   ) ^{1/2} .
\label{eq:phccdisp}
\end{eqnarray}
We note that  simpler dispersion relations of the form
$ E({\bf Q}) = A_{0} + \sum_{i}A_{i}\cos({\bf Q}\cdot {\bf R}_{i})$,
which provide an effective description of related systems
with weaker dispersion such as CuHpCl, \cite{Stone01} do not provide
a good description of the dispersion in PHCC.
Because most of the observed magnetic intensity comes in the form of 
resolution-limited peaks, we can use the Single Mode Approximation 
(SMA)
for the dynamic correlation function,
\begin{equation}
{\cal S}^{\alpha \alpha}({\bf Q},\omega ) = {\cal S}({\bf Q})
\delta(\hbar\omega - E({\bf Q})) ,
\label{eq:sma1}
\end{equation}
for  quantitative analysis of the data. Here ${\cal S} ({\bf Q})$ is 
the static
structure factor, which is the Fourier transform of the equal-time
two-spin correlation function, and quantifies the energy-integrated
intensity (it is the zeroth moment of ${\cal S}^{\alpha\alpha}({\bf
Q},\omega)$). The polarization index $\alpha$
may be omitted for an isotropic spectrum.
The SMA has been used successfully for
gapped spin chains,\cite{gxu2000,ma92} and other quantum many-body
systems  \cite{girvin86}
where a single, coherent mode dominates the excitation spectrum.
Using the SMA expression (Eq.~\ref{eq:sma1}) in the first moment sum
rule (Eq.~\ref{eq:firstmomsumr}) one can establish
a relation between the measured
intensities and the dispersion relation E(Q),
\begin{equation}
\label{SqSMA}
 {\cal S} ({\bf Q}) = -\frac{1}{3E({\bf Q})}\sum_{\bf d} J_{\bf d}
\langle {\bf S_0} \cdot {\bf S_d}\rangle\left( 1-\cos{\bf Q\cdot
d}\right) \;,
\end{equation}
that involves the spin pair correlations which contribute to the 
ground state
energy.

We have carried out  global fits to
$\tilde{I}_m({\bf Q},\hbar \omega)$ for our entire
data set, using Eqs.~\ref{eq:phccdisp}-\ref{SqSMA} as input
to Eq.~\ref{eq:ItildeQw}.  For each energy scan,
the non-magnetic background was modeled
by a constant plus a Gaussian peak centered at $\hbar \omega = 0$
to account for incoherent elastic scattering as needed.
To determine the simplest parametrization of the data, global fits
were carried out with varying numbers of
terms included in  Eq.~\ref{SqSMA}.  The
results of
these fits with  Bonds 1-6 and Bonds 1-8 in Fig.~\ref{fig:phccstruct}
are given in Tables  \ref{tab:phcctable} and \ref{tab:phccdisptable}.
Further neighbor bonds gave no measurable contribution when included 
in
the fits.
Examples of the lineshapes derived from the 8-bond
fit are shown as solid lines
in Figs.~\ref{fig:escansH}-\ref{fig:escansL}.  
The right-hand
vertical scale in these figures gives the normalized magnetic 
intensity
$\tilde{I}_m({\bf Q},\hbar\omega )$. Magnetic neutron scattering from 
a local moment magnet satisfies a total moment sum rule that can be 
expressed as
\begin{equation}
\langle S^2 \rangle = \frac{\hbar\int d^3{\bf Q} 
d\omega\sum_{\alpha\beta}
{\cal S}^{\alpha\beta}({\bf Q},\omega )}{\int d^3{\bf Q} }=S(S+1)
\end{equation}
Clearly we do not have enough data to carry out the complete 
integration. However, we can ask whether the SMA, which accounts for 
the scattering data where available, satisfies the sum rule. Carrying 
out the $\bf Q$-integration of Eq.~\ref{SqSMA} numerically, we obtain 
$\langle S^2\rangle =0.8(2)$. The result being indistinguishable from 
$S(S+1)=3/4$, suggests that the resonant mode accounted for by the 
SMA carries most of the spectral weight in PHCC. 

The dispersion relation determined from the 8-bond global fit is shown
as solid lines in Figs.~\ref{fig:phccdispH}-\ref{fig:phccdispL}.  It
is consistent with the data derived directly from the raw data,
indicating that the variational dispersion relation
Eq.~\ref{eq:phccdisp} used in the SMA is general enough not to
cause significant bias.  As may be seen from
Table~\ref{tab:phccdisptable}, the parameters in Eq.~\ref{eq:phccdisp}
are well determined, and the results for the 6- and 8-bond fits are
indistinguishable.

The first moment calculated from the model with
the parameters determined from the global fits is shown
in Figs.~\ref{fig:fmh}-\ref{fig:fml}
as dashed and solid lines
for the 6-bond and 8-bond fits, respectively.
Instrumental resolution effects were included in this calculation.
The agreement with the first moment determined from the
Gaussian fits to individual scans is in general quite good,
with modest improvement in the quality of the 8-bond fit over
the 6-bond fit.   There is also good quantitative agreement
for the terms common to both fits, particularly
for the larger terms, indicating the robustness of the
numbers determined through this analysis.

Without a microscopic model that connects exchange constants with the
dispersion relation, it is not possible to determine $J_{\bf d}$ and
$\langle{\bf S}_0\cdot{\bf S}_{\bf d}\rangle$ independently.
Instead one measures their products, which determine the contribution
of each bond to the ground state energy $E_{0}$ [Eq.~\ref{eq:e0}].
In fact, because $|\langle{\bf S}_0\cdot{\bf S}_{\bf
d}\rangle|<3/4$, each term provides a lower bound on the magnitude of
the corresponding exchange interaction.  In 
Fig.~\ref{fig:phccstruct}(c),
the thickness of the Cu-Cu bonds is proportional to
$|J_{{\bf d}}\langle{\bf S}_0\cdot{\bf S}_{\bf d}\rangle|$. From
this figure and from Table
\ref{tab:phcctable}, we see that the largest Cu-Cu interaction is Bond
1, the ``dimer'' predicted previously.  \cite{daoud86} However, other
bonds also show significant antiferromagnetic correlations,
particularly Bonds 3 and 6.
These bonds are noteworthy as they provide uniform albeit 
anisotropic, linkage of a macroscopic two dimensional spin system. It 
is also important to note that several
bonds give a positive contribution to the first moment. These bonds,
shown as gray lines in Fig.~\ref{fig:phccstruct}(c), are
frustrated by definition as they raise the ground state energy.

Summing up all terms shown in Table \ref{tab:phcctable}
we find the ground state energy
per spin $E_0 = -1.5(4)$ meV  for both the  6- and 8-bond
fits. Using the measured value of spin gap $\Delta = 1$ meV, we can
thus determine the relation $E_0 \approx -1.5\cdot \Delta$, which can
be used to quantify explicitly the correspondence of the spin system 
of
PHCC to a particular quantum-disordered 2D spin model. It also helps 
to
position PHCC on the phase diagram of the 2D QHAFM 
\cite{Chakravarty1988},
characterizing how close it is to quantum criticality.

\section{Conclusions}

The set of measurements presented here establish PHCC as a clean 
example
of a two-dimensional spin-singlet system with a gap. The coupled-dimer
picture that describes previously studied 2D and 3D spin gap materials
is not appropriate for PHCC. This material lies in an unusual region 
of
the phase space of the model, in a quantum disordered regime close to
the quantum-critical point. Moreover, the susceptibility and heat 
capacity
measurements show evidence of a quantum phase transition from the
quantum disordered to
the N\'{e}el-ordered renormalized-classical state at a magnetic field
$H_{c1} \approx 7.5$ T. Although we are not yet able to associate
unambiguously the zero-field ground state with either a particular 
kind
of valence bond crystal
or a RVB-type spin-liquid, it is clear that frustration plays
a key role in defining it. It is worth noting that evidence for
spin-frustration effects was recently observed in CuHpCl
\cite{Stone01}, another quasi-2D spin system, as well as in
SrCu$_2$(BO$_3$)$_2$ \cite{kageyama99}, and thus it may be that
frustration is in fact a common feature of low-symmetry, gapped spin
systems.

Another issue that becomes increasingly important as one moves towards
the QC point, or from the dimerized limit toward a system with more
uniform coupling, is the appearance of a continuum in the spin
fluctuation spectrum. In the quantum-disordered 2D HAFM models this is
described in terms of spinon deconfinement \cite{Chubukov1995}, while 
in
the strongly dimerized limit it is usually adequately accounted for by
multi-magnon excitations.  There are some features in the present data
indicating that a continuum may be detectable in
PHCC [e.g. the broad, weak feature at
$\hbar\omega \approx 3.5$ meV in Fig.~\ref{fig:escansK}], but further 
experiments are required to clarify this issue.
Finally, the low energy scales in PHCC make it an excellent candidate
for further studies of quantum critical and field-dependent
phenomena in a two-dimensional gapped spin system with frustration.

\section{Acknowledgments}
We thank R. Paul for help with neutron activation analysis.
This work was supported by  NSF Grant DMR-9801742.  DHR
acknowledges the support of the David and Lucile Packard Foundation.
X-ray characterization and
SQUID magnetometry was carried out using facilities maintained by
the JHU MRSEC under NSF Grant number DMR-0080031.  This work utilized
neutron research facilities supported by NIST and the NSF under
Agreement No. DMR-9986442.

\begin{table}
\begin{tabular}{crrrrrr}
            &      &   &   &
                & 6 Bonds
                & 8 Bonds
                \\
Bond number & $x/a$ & $y/b$ & $z/c$    & $|{\bf d}|$ (\AA)
				& \multicolumn{1}{c}{$J_{\vec{d}}\langle{\bf
S}_0\cdot{\bf S}_{\vec{d}}\rangle$}
				& \multicolumn{1}{c}{$J_{\vec{d}}\langle{\bf
S}_0\cdot{\bf S}_{\vec{d}}\rangle$}
              \\  \hline
1   & -0.19 &  0.12 & 0.51   & 3.450  & -1.3(3) &  -1.4(3)   \\
2   &  0.19 & -0.12 & 0.49   & 3.442  &  0.7(3) &   0.6(3)  \\
3   &  0    &  0    & 1      & 6.104  & -0.3(1) &  -0.4(1)  \\
4   &  0.81 &  0.12 & 0.51   & 6.730  &  0.1(3) &  -0.2(3)  \\
5   &  0.81 &  0.12 &-0.49   & 7.879  & -0.0(3) &  -0.1(3)  \\
6   &  1    &  0    & 0      & 7.984  & -0.92(5) &  -0.95(5)  \\
7   &  1.19 & -0.12 & 0.49   & 9.439  &          &   0.1(2)  \\
8   & -1.19 &  0.12 & 0.51   &10.296  &          &   0.6(2)

\end{tabular}
\caption{Fractional coordinates (from Ref. [20]),
bond lengths, and corresponding
values of $J_{{\bf
d}}\langle{\bf
S}_{0}\cdot{\bf S}_{{\bf d}} \rangle$ (in meV) for the
fits of the PHCC inelastic neutron scattering data including 6
and 8 Cu-Cu bonds.  Bond
numbers correspond to those depicted in Fig.~\ref{fig:phccstruct}(c).}
\label{tab:phcctable}
\end{table}

\begin{table}
\begin{tabular}{lll}
 \multicolumn{1}{c} Parameter &  \multicolumn{1}{c}{6 Bonds
(meV$^{2}$)}  &
\multicolumn{1}{c}{8
          Bonds (meV$^{2}$)}
                \\ \hline
$B_{0}$   &  5.44(2)       & 5.45(2)   \\
$B_{h}$   &  2.06(3)       & 2.05(3)  \\
$B_{l}$   &  1.07(3)       & 1.06(3)   \\
$B_{hl}$  & -0.39(1)      & -0.39(1) \\
$B_{2h}$  & -0.34(3)       & -0.35(3)  \\
$B_{2l}$  & -0.22(2)       & -0.23(2)  \\

\end{tabular}
\caption{Fitted  parameters in the dispersion
relation, Eq.~\ref{eq:phccdisp}, for the 6- and 8-bond fits
of the PHCC inelastic neutron scattering data. }
\label{tab:phccdisptable}
\end{table}

\newpage

\end{document}